\documentclass[preprint,aps,pra,showpacs]{revtex4}
\usepackage{graphicx}
\usepackage{amsmath}
\usepackage{amsfonts}
\usepackage{bm}
\usepackage{color}

\def\ketm#1{  \left\vert  #1   \right\rangle   }

\def\mem#1#2#3{  \left\langle #1 \left\vert  #2 \right\vert #3 \right\rangle   }

\newcommand{\balpha}{{\mbox{\boldmath$\alpha$}}}

\newcommand{\be}{\begin{eqnarray}}

\newcommand{\ee}{\end{eqnarray}}

\newcommand{\la}{\langle}

\newcommand{\ra}{\rangle}

\newcommand{\bfx}{{\bf x}}

\newcommand{\bfk}{{\bf k}}

\newcommand{\bfp}{{\bf p}}

\newcommand{\eps}{\epsilon}

\newcommand{\beps}{{\mbox{\boldmath$\epsilon$}}}

\begin{document}

\title{Parity nonconservation in radiative recombination of electrons with
heavy hydrogenlike ions}

\author{A.V. Maiorova,$^{1}$ O.I. Pavlova,$^{1}$ V. M. Shabaev,$^{1}$
C. Kozhuharov,$^{2}$ G. Plunien,$^{3}$  and T. St\"ohlker$^{2}$}
\address{$^{1}$ Department of Physics, St.Petersburg State University,
Ulianovskaya 1, Petrodvorets, St.Petersburg 198504, Russia}
\address{$^{2}$ Gesellschaft f\"{u}r Schwerionenforschung,
Planckstrasse 1, D-64291 Darmstadt, Germany}
\address{$^3$ Institut f\"ur Theoretische Physik, TU Dresden,
Mommsenstra{\ss}e 13, D-01062 Dresden, Germany}

\begin{abstract}

The parity nonconservation effect on the radiative recombination
of electrons with heavy hydrogenlike ions is studied. Calculations
are performed for the recombination into the $2^1S_0$ state of
helium-like thorium and gadolinium, where, due to the
near-degeneracy of the opposite-parity $2^1S_0$ and  $2^3P_0$
states, the effect is strongly enhanced. Two scenarios for
possible experiments are studied. In the first scenario, the
electron beam is assumed to be fully polarized while the H-like
ions are unpolarized and the polarization of the emitted photons
is not detected. In the second scenario, the linearly polarized
photons are detected in an experiment with unpolarized electrons
and ions. Corresponding calculations for the recombination into
the  $2^3P_0$ state are presented as well.

\end{abstract}


\maketitle

\section{Introduction}

Investigations of the parity nonconservation (PNC) effects in
atoms remain an effective tool for tests of the standard model
(SM) and its various extensions \cite{khr91,khr04,gin04}.
High-precision measurement of the 6$s$-7$s$ PNC amplitude in
$^{133}$Cs  \cite{wood97,ben99}, combined with the recent progress
on the QED and atomic-structure calculations
\cite{der00,koz01,sush01,joh01,dzu02,kuch02,mil02,sap03,sha05,por09},
provided the most accurate to-date test of the electroweak  sector
of the SM at the low-energy regime. From the theoretical side, one
of the main difficulties in calculations of the PNC effects in
neutral atoms consists in the high-precision evaluation of the
electron-correlation contributions (see Ref. \cite{por09} and
references therein). This problem disappears if one deals with
few-electron highly charged ions, where the electron-correlation
effects, being suppressed  by a factor $1/Z$ ($Z$ is the nuclear
charge number), can be evaluated by perturbation theory to the
required accuracy.

The PNC effects in highly charged ions were first discussed by
Gorshkov and Labzowsky in Refs.  \cite{gor74}, where a proposal to
use close opposite-parity levels  $2^1S_0$ and  $2^3P_1$ for
$Z\approx 6$ and  $Z\approx 29$ was made. An idea for detecting
parity violation in He-like ions with $Z\approx 6$ by
investigating the induced $2^3S_1$ - $2^1S_0$ transition in the
presence of electric and magnetic fields was considered by von
Oppen \cite{opp91}. Various scenarios for observing the PNC effect
in He-like uranium using the near-degeneracy of the $2^1S_0$ and
$2^3P_0$ states were discussed in Refs. \cite{sch89,kar92,dun96}.
Sch\"afer {\it et al.} \cite{sch89} estimated the laser
intensities required to observe the PNC asymmetry in the
two-photon $2^3P_0$ -  $2^1S_0$ transition. Karasiev {\it et al.}
\cite{kar92} evaluated the degree of circular polarization of
photons emitted in the hyperfine-quenched  one-photon  $2^1S_0$ -
$1^1S_0$ transition. An idea to study the PNC effect on the
two-photon $2^3P_0$ -  $1^1S_0$ transition, stimulated by the
circularly polarized optical laser, was proposed by Dunford
\cite{dun96}. PNC experiments with polarized ion beams at
$Z\approx 64$, where  the $2^1S_0$ and  $2^3P_0$ states of He-like
ions are also near degenerate, were suggested by Labzowsky {\it et
al.} \cite{lab01}. As in Ref. \cite{kar92}, here the
hyperfine-induced one-photon  $2^1S_0$ - $1^1S_0$ transition was
considered. A detailed analysis of possibilities for the PNC
experiments with heavy H-like ions was presented by Zolotarev and
Budker \cite{zol97}. The parity-violating effect on the Auger
decay of doubly-excited states of He-like uranium was examined by
Pindzola \cite{pin93}. In Ref.  \cite{gri05}, Gribakin {\it et
al.} have studied the PNC effect on the cross section of
dielectronic recombination into doubly excited states of He-like
ions at $Z<60$.

In the present paper, we study the PNC effect on the one-photon
radiative recombination (RR) of an electron into the $2^1S_0$ and
the $2^3P_0$ state of He-like ions nearby $Z=90$ and $Z=64$, where
the opposite-parity states  $2^1S_0$ and  $2^3P_0$ are close to
crossing.

Relativistic units ($\hbar=c=1$) and the Heaviside charge unit
($\alpha = e^2/(4\pi)$, $e<0$) are used throughout the paper.

\section{Basic formulas}

Theory of the radiative recombination of electrons with
highly-charged ions was considered by many authors
\cite{eichler_book_95,sha02,kla02,sur02,fritzsche05,eic07,mai09}.
In the present paper, we consider the one-photon radiative
recombination of an electron having the asymptotic four-momentum
$p_i = (p_i^0,\bfp)$ and the spin projection $\mu_i$ with a heavy
H-like ion being originally in the $1s$ ground state. Here --- and
in what follows
--- it is assumed  that the momentum   $\bfp_i$ is directed along
the quantization  axis ($z$ - axis). Since we are interested in
the PNC effect, we consider that the electron is captured into the
$2^1S_0$ (or, alternatively, $2^3P_0$)  state of a heavy
helium-like ion with $Z\approx 90$ or $Z\approx 64$, where the
opposite-parity states $2^1S_0$ and $2^3P_0$ are near degenerate.
This capture is  accompanied by the emission of a photon with
momentum $\bfk$, energy $k^0=|\bfk|= p_i^0 - E_{2^1S_0}$, and
polarization  $\eps^{\nu}=(0,\beps)$. To zeroth order, the cross
section of the process is given by \cite{sha02}
\begin{eqnarray}
\frac{d\sigma}{d\Omega} =\frac{(2\pi)^4}{v_i}\bfk^2|\la
f|R^{\dag}(1) + R^{\dag}(2)|i\ra|^2 \,, \label{cross}
\end{eqnarray}
where $|i\ra$ and $|f\ra$ denote the initial and final states of
the two-electron system, $R =-e\balpha\cdot {\bf A}$ is the
transition operator acting on the electron variables labeled in
Eq. (\ref{cross}) by the indices 1 and 2, respectively,
\begin{eqnarray}
{\bf A}(\bfx) = \frac{\beps \exp{(i\bfk\cdot
\bfx)}}{\sqrt{2k^0(2\pi)^3}} \, \label{photon}
\end{eqnarray}
is the wave function of the emitted photon, and $v_i$ is the
initial electron velocity. Since for heavy few-electron ions the
interelectronic-interaction effects are suppressed by a factor
$1/Z$, compared to the electron-nucleus Coulomb interaction, we
can consider the wave functions of the initial and final states in
the one-electron approximation. The uncertainty due to neglecting
the interelectronic-interaction and  QED corrections should  not
exceed  a few-percent level \cite{yer00,sur08,sha00}. With this
approximation, the initial state is described by the wave function
\begin{eqnarray}
u_i(\bfx_1,\bfx_2) = \frac{1}{\sqrt{2}}(\psi_{j\,m}(\bfx_1)
\psi_{p_i\,\mu_i}(\bfx_2)-\psi_{j\,m}(\bfx_2)
\psi_{p_i\,\mu_i}(\bfx_1))\,, \label{wf_i}
\end{eqnarray}
where $\psi_{j\,m}(\bfx)$ is the one-electron $1s$ wave function
and  $\psi_{p_i\,\mu_i}(\bfx)$ is the incident electron wave
function. If we neglect the interelectronic and the weak
electron-nucleus interaction, the wave function of the final
($2^1S_0$ or $2^3P_0$) state is given by
\begin{eqnarray}
u_f(\bfx_1,\bfx_2) = \frac{1}{\sqrt{2}}\sum_{m_1,m_2}
C^{00}_{j_1\,m_1,j_2\,m_2}
(\psi_{j_1\,m_1}(\bfx_1)\psi_{j_2\,m_2}(\bfx_2)
-\psi_{j_1\,m_1}(\bfx_2)\psi_{j_2\,m_2}(\bfx_1)) \,, \label{wf_f}
\end{eqnarray}
where $\psi_{j_1\,m_1}(\bfx)$ is the one-electron $1s$ wave
function, $\psi_{j_2\,m_2}(\bfx)$ is the one-electron $2s$
($2p_{1/2}$) wave function, and  $C^{JM}_{j_1\,m_1,j_2\,m_2}$ is
the Clebsch-Gordan coefficient. To account for the weak
interaction we must modify the wave function of the $2^1S_0$
($2^3P_0$) state by admixing the $2^3P_0$ ($2^1S_0$) state. This
yields
\begin{eqnarray}
\label{wf_pnc1} |2^1S_0\ra \rightarrow |2^1S_0\ra +\frac{\la
2^3P_0|H_W(1)+H_W(2)|2^1S_0\ra}{E_{2^1S_0}-
E_{2^3P_0}}|2^3P_0\ra\,,\\
|2^3P_0\ra \rightarrow |2^3P_0\ra +\frac{\la
2^1S_0|H_W(1)+H_W(2)|2^3P_0\ra}{E_{2^3P_0}-
E_{2^1S_0}}|2^1S_0\ra\,, \label{wf_pnc2}
\end{eqnarray}
where \be \label{h_w} H_W=-(G_F/\sqrt{8})Q_W \rho_{N}(r)\gamma_5
\ee is the nuclear spin-independent weak-interaction Hamiltonian
\cite{khr91}, $G_F$ is the Fermi constant,  $Q_W \approx -N +
Z(1-4{\rm sin}^2\theta_W) $ is the weak charge of the nucleus,
$\gamma_5$ is the Dirac matrix, and $\rho_{N}$  is the nuclear
weak-charge density normalized to unity. A simple evaluation of
the weak-interaction matrix element gives
\begin{eqnarray} \label{h_w_m} \la
2^3P_0|H_W(1)+H_W(2)|2^1S_0\ra&=& \la 2p_{1/2}|H_W|2s\ra
\nonumber\\
 &=& i\frac{G_F}{2\sqrt{2}}Q_W\int_{0}^{\infty}dr\,r^2
\rho_N(r)[g_{2p_{1/2}}f_{2s}-f_{2p_{1/2}}g_{2s}]\,.
\end{eqnarray}
The large and small radial components of the Dirac wave function,
$g(r)$ and $f(r)$, are defined by
\begin{eqnarray} \psi_{n\kappa m}({\bf r})=
\left(\begin{array}{c}
g_{n\kappa}(r)\Omega_{\kappa m}({\bf n})\\
if_{n\kappa}(r)\Omega_{-\kappa m}({\bf n})
\end{array}\right)\;,
\end{eqnarray}
where $\kappa=(-1)^{j+l+1/2}(j+1/2)$ is the Dirac quantum number.
Then formulas (\ref{wf_pnc1})-(\ref{wf_pnc2}) can be written as
\begin{eqnarray}
\label{wf_pnc1p}
|2^1S_0\ra \rightarrow |2^1S_0\ra +i\xi|2^3P_0\ra\,,\\
|2^3P_0\ra \rightarrow |2^3P_0\ra +i\xi|2^1S_0\ra\,,
\label{wf_pnc2p}
\end{eqnarray}
where \be \label{xi} \xi =
\frac{G_F}{2\sqrt{2}}\frac{Q_W}{E_{2^1S_0}-E_{2^3P_0}}
\int_{0}^{\infty}dr\,r^2
\rho_N(r)[g_{2p_{1/2}}f_{2s}-f_{2p_{1/2}}g_{2s}]\,. \ee With this
correction,  the differential cross section of recombination into
the $2^1S_0$ state, $\sigma \equiv d\sigma/d\Omega $, can be
written in terms of the one-electron matrix elements:
\begin{eqnarray}
\sigma = \frac{1}{2}\frac{(2\pi)^4}{v_i}\bfk^2 \Bigl\{|\la 2s
-m|R^{\dag}|p_i\mu_i\ra|^2 +2\Re{[i\xi\la 2s
-m|R^{\dag}|p_i\mu_i\ra \la p_i \mu_i|R|2p_{1/2} -m\ra]} \Bigr\}
\,, \label{cross1}
\end{eqnarray}
where $m$ is the angular momentum projection of the initial $1s$
electron. The corresponding expression for the recombination into
the $2^3P_0$ state is given by
\begin{eqnarray}
\sigma = \frac{1}{2}\frac{(2\pi)^4}{v_i}\bfk^2 \Bigl\{|\la
2p_{1/2} -m|R^{\dag}|p_i\mu_i\ra|^2 +2\Re{[i\xi\la 2p_{1/2}
-m|R^{\dag}|p_i\mu_i\ra \la p_i \mu_i|R|2s -m\ra]} \Bigr\} \,.
\label{cross2}
\end{eqnarray}
The incoming electron wave function is given by the partial wave
expansion
\begin{eqnarray}
\label{electron_wave_decompose} \ketm{p_{i}\mu_{i}} =
\frac{1}{\sqrt{4\pi}} \, \frac{1}{\sqrt{p_{i}\varepsilon_{i}}} \,
\sum_{\kappa} \mathrm{i}^{l}\exp(i\Delta_{\kappa}) \, \sqrt{2l+1}
\,
 C_{l0, \,\frac{1}{2}\mu_{i}}^{j\mu_{i}} \,
\ketm{\varepsilon_{i} \kappa \mu_{i}} \, ,
\end{eqnarray}
where $\Delta_{\kappa}$ is the Coulomb phase shift and
$\ketm{\varepsilon_{i} \kappa \mu_{i}}$ is the partial electron
wave with the energy $\varepsilon_{i}=p^{0}_{i}$ and the Dirac
quantum number $\kappa$. This expansion enables one to express the
free--bound transition amplitude as a sum of partial amplitudes
\begin{eqnarray}
\mem{p_{i}\mu_{i}}{R}{n_b j_b \mu_b} &=& \frac{1}{\sqrt{4\pi}} \,
\frac{1}{\sqrt{p_{i}\varepsilon_{i}}} \, \sum_{\kappa}
(\mathrm{-i})^{l} \, \exp(-i \Delta_{\kappa}) \, \sqrt{2l+1}
\nonumber \\
&\times& C_{l0, \,\frac{1}{2}\mu_{i}}^{j\mu_{i}} \,
\mem{\varepsilon_{i} \kappa \mu_{i}}{R}{n_b j_b \mu_b} \, .
\end{eqnarray}
The latter amplitude is evaluated employing the standard partial
wave decomposition for the photon wave function (see, e.g., Refs.
\cite{eic07,ros57}). The angular integrations are carried out
analytically while the radial integrations are accomplished
numerically. The RADIAL package \cite{salvat} is used to calculate
the bound and continuum wave functions for extended nuclei.

\section{Results and discussion}

The formulas (\ref{cross1})-(\ref{cross2}) represent the
differential cross section at given values of the bound-electron
angular momentum projection $m$, the incoming electron
polarization  $\mu_i$, and the outgoing photon polarization
$\beps$. To investigate the role of the PNC effect, we consider
two different scenarios for an experiment. In the first scenario,
the incident electron is polarized, while the H-like ion is
unpolarized, and the photon polarization is not detected. In this
scenario, the cross sections  (\ref{cross1})-(\ref{cross2}) must
be averaged over  the bound-electron angular momentum projection
$m$ and summed over the outgoing photon polarization $\beps$.
Since recent advances in polarization techniques
\cite{Stohlker03,Tashenov06} make measurements of the linear
polarization of x--rays feasible, as the second one we consider a
scenario, in which linearly polarized photons are detected in an
experiment with unpolarized electrons and ions. In this scenario,
we have to avarage over $m$ and $\mu_i$, and consider the photon
linearly polarized under the angle $\chi $ with respect to the
reference plane that is spanned by the incident electron and the
emitted photon momenta (Fig. 1).

%
% ------------------------- Figure 1 ---------------------------- %
%
\begin{figure}
\includegraphics[width=\columnwidth]{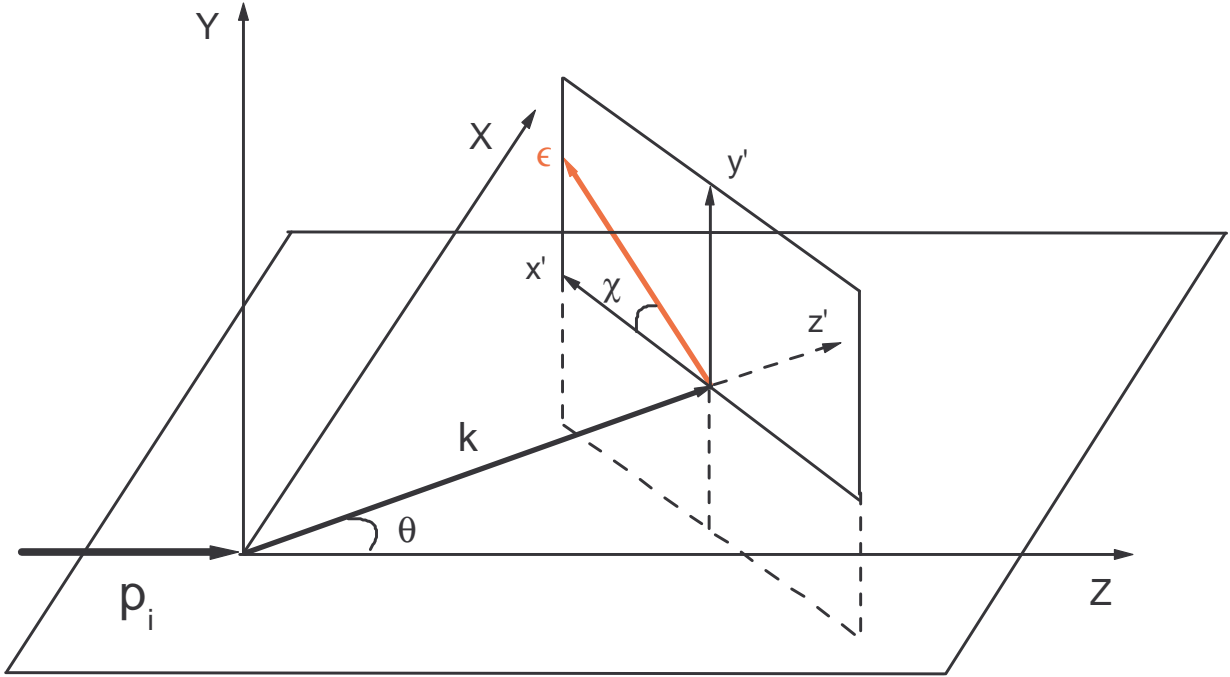} \caption{Geometry (in the ion rest frame) for
one-photon radiative recombination of a free electron into an
excited state of a projectile ion. The unit vector of the linear
polarization of the photon is defined in the plane that is
perpendicular to the photon  momentum.} \label{Fig1}
\end{figure}

Since we are interested in observing the PNC effects, we should
search for situations where these effects are enhanced as much as
possible. As indicated above, the most promising situation occurs
in cases where the $2^1S_0$ and  $2^3P_0$ levels almost cross. The
ions with $Z\approx 90$ and  $Z\approx 64$ are presently
considered as the best candidates for that. In Table I, we list
the theoretical predictions for the $2^3P_0$ -  $2^1S_0$ energy
difference in ions near the crossing points  \cite{art05,kozh08a}.
Compared to Ref. \cite{art05}, these data are obtained using the
revised values of the nuclear charge radii for $^{238}$U and
$^{232}$Th \cite{kozh08a,kozh08b} as well as recent results for
the two-loop QED contributions \cite{yer06}. As seen from the
table, the maximum enhancement takes place in the cases of Th and
Gd. Although the current theoretical accuracy is not high enough,
one can expect that the energy differences considered can be
determined to the desired accuracy in an experiment \cite{bra08}.
In accordance with the table, to estimate the PNC effect we  use
0.44 eV and 0.074 eV for the   $2^3P_0$ -  $2^1S_0$  energy
difference in the cases of Th and Gd, respectively.  We note that
in both cases the energy differences utilized are significantly
larger than the corresponding natural line widths.

As the next step, one should determine the sensitivity
requirements for an experimental apparatus capable for observing
the PNC effect to a given accuracy. Let us consider these
requirements for the first experimental scenario with a fully
polarized electron beam. Denoting by $\sigma_{+}(\theta)$ and
$\sigma_{-}(\theta)$ the cross sections for the positive and
negative helicities (the spin projection onto the electron
momentum direction) of the incident electron, we can write for the
related numbers of counts
\begin{eqnarray}
N_{\pm}=LT(\sigma_{\pm} + \sigma_{\rm b}) \,,
\end{eqnarray}
where $\sigma_{\rm b}$ is the background magnitude, $T$ is the
acquisition time, and $L$ is the luminosity defined by the
experimental conditions. Let us assume that we want to measure the
PNC effect with a relative uncertainty $\eta$. Then, taking into
account that the statistical error of $N_{+}-N_{-}$ is given by
$\sqrt{N_{+}+N_{-}}$, one derives the following requirement for
the luminosity (cf. Ref. \cite{gri05})
\begin{eqnarray}
L>L_{0}=\frac{\sigma_+ + \sigma_-+2\sigma_{\rm
b}}{(\sigma_+-\sigma_-)^2\eta^2 T} \,. \label{lum}
\end{eqnarray}
For the following analysis we neglect the background signal
$\sigma_b$ and assume the acquisition time $T$ is equal two weeks.

We calculated $L_{0}$ for different inclination angles $\theta$
and different incident electron energies. Table II presents
numerical results for the radiative recombination into the
$2^1S_0$ and the $2^3P_0$ state of He-like thorium at  the angles
$\theta$ corresponding to the minimum values of the luminosity
$L_0$. For completeness, the cross section without the PNC effect,
$\sigma_0 = (\sigma_+ +\sigma_-)/2$, and the PNC contribution,
$\sigma_{\rm PNC} = (\sigma_+ -\sigma_-)/2$, are presented as
well. In Figs. 2 and 3, we display the values $\sigma_{\rm
PNC}^2/\sigma_0 \sim 1/L_0$ as functions of $\theta$ for the
radiative recombination into the  $2^1S_0$ and the $2^3P_0$ state,
respectively, at the incident electron energy of 1 eV.

%
% ------------------------- Figure 2 ---------------------------- %
%
\begin{figure}
\includegraphics[width=\columnwidth]{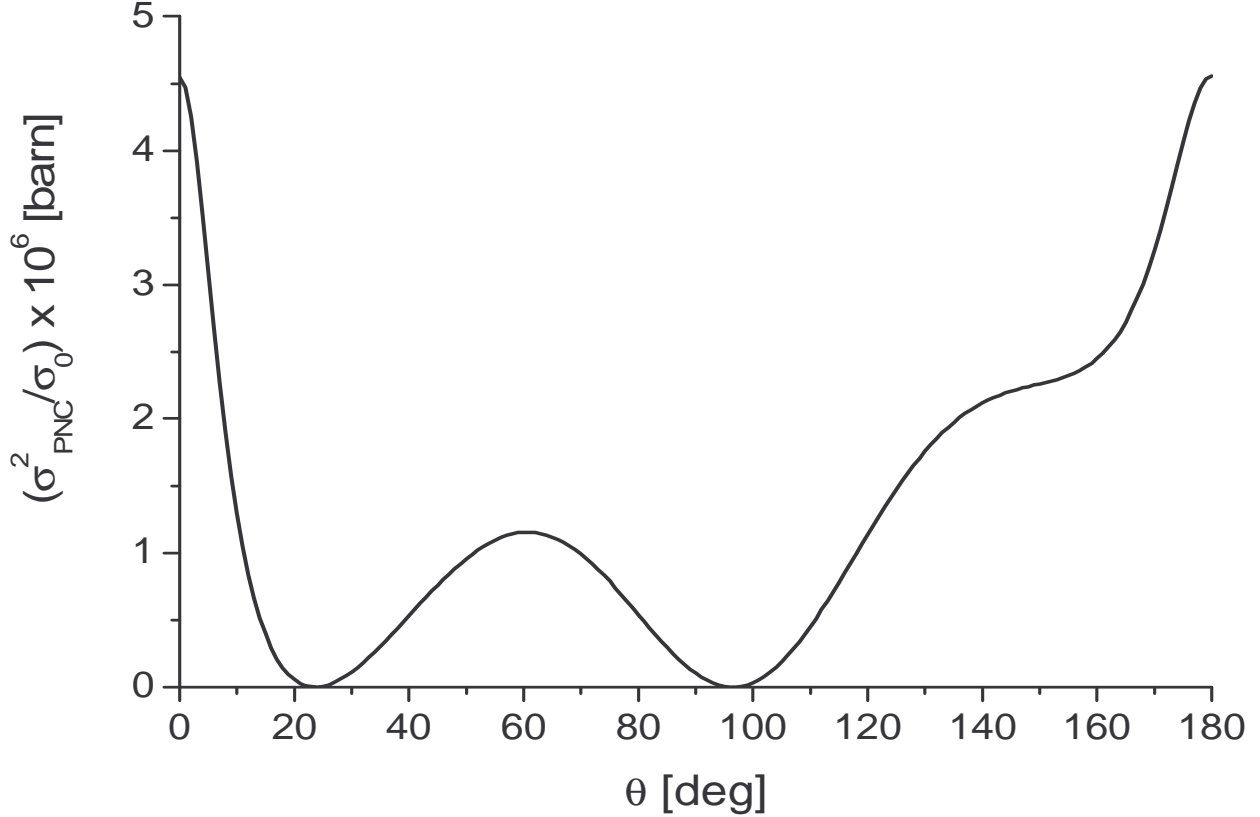} \caption{ The value $\sigma_{\rm
PNC}^2/\sigma_0 \sim 1/L_0$ as a function of $\theta$ for the
radiative recombination into the  $2^1S_0$ state of He-like
thorium at the incident electron energy of 1 eV. It is assumed
that the incoming electron is polarized while the H-like ion is
unpolarized and the photon polarization is not detected. }
\label{Fig2}
\end{figure}

%
% ------------------------- Figure 3 ---------------------------- %
%
\begin{figure}
\includegraphics[width=\columnwidth]{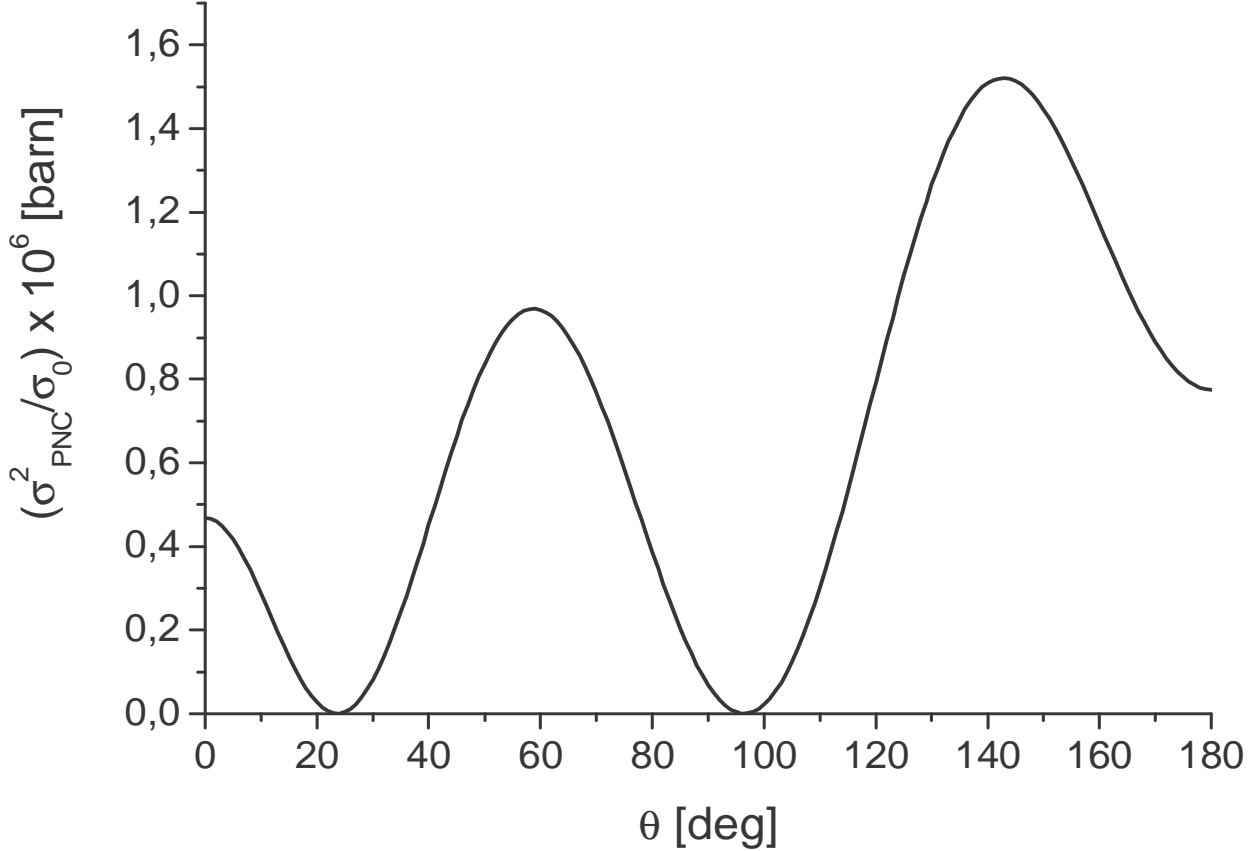} \caption{ The value $\sigma_{\rm
PNC}^2/\sigma_0 \sim 1/L_0$ as a function of $\theta$ for the
radiative recombination into the  $2^3P_0$ state  of He-like
thorium at the incident electron energy of 1 eV. It is assumed
that the incoming electron is polarized while the H-like ion is
unpolarized and the photon polarization is not detected. }
\label{Fig3}
\end{figure}

Tables III and IV present numerical results for the second
scenario, where linearly polarized photons are detected in an
experiment with unpolarized electrons and ions. As before,
$\sigma_0$ denotes the cross section without the PNC effect and
$\sigma_{\rm PNC}$ is the PNC contribution. Again,  the angles
$\theta$ and $\chi$ considered in Tables III and IV correspond to
the minimum values of the luminosity. In this scenario, the sign
of the PNC contribution $\sigma_{\rm PNC}$ is changed if the
polarization angle $\chi$ is replaced by $\pi-\chi$. In Figs. 4
and 5, we display the value $\sigma_{\rm PNC}^2/\sigma_0 \sim
1/L_0$ as a function of $\theta$  and $\chi$ for the RR into the
$2^1S_0$ and the $2^3P_0$  state, respectively, at the incident
electron energy of 1 eV. The change of the sign of the PNC
contribution $\sigma_{\rm PNC}$ under the replacement $\chi
\rightarrow \pi-\chi$ means, in particular, that measuring the
count rate difference between the two linear polarizations, which
can be achieved in the same experiment by setting the detectors at
different azimuth angles $\phi$, can provide a direct access to
the pure PNC effect. With this in mind, in Table V we present
numerical results for the angles   $\theta$ and $\chi$
corresponding to the maximum  absolute values of   $\sigma_{\rm
PNC}$. We note that the contribution  $\sigma_{\rm PNC}$ has the
same absolute values but   carries opposite  signs for the
radiative recombination into the  $2^1S_0$ and the  $2^3P_0$
state, respectively. It follows that the PNC effect disappears if
one measures the cross section of the RR into both  $2^1S_0$ and
$2^3P_0$ states. To observe the PNC effect we need to detect the
photons that originate from only one of these processes: either
from the RR into the $2^1S_0$ state or from the RR into the
$2^3P_0$ state.  Although at present the experimental resolution
is far from being sufficient to detect  the desired transition
line, we think that with some experimental ingenuity the blinding
line could be eliminated. Alternatively,  one might consider the
same experiment at other values of $Z$, where the $2^3P_0$ -
$2^1S_0$ energy difference becomes larger while the PNC effect
remains still sizeable.

%
% ------------------------- Figure 4 ---------------------------- %
%
\begin{figure}
\includegraphics[width=\columnwidth]{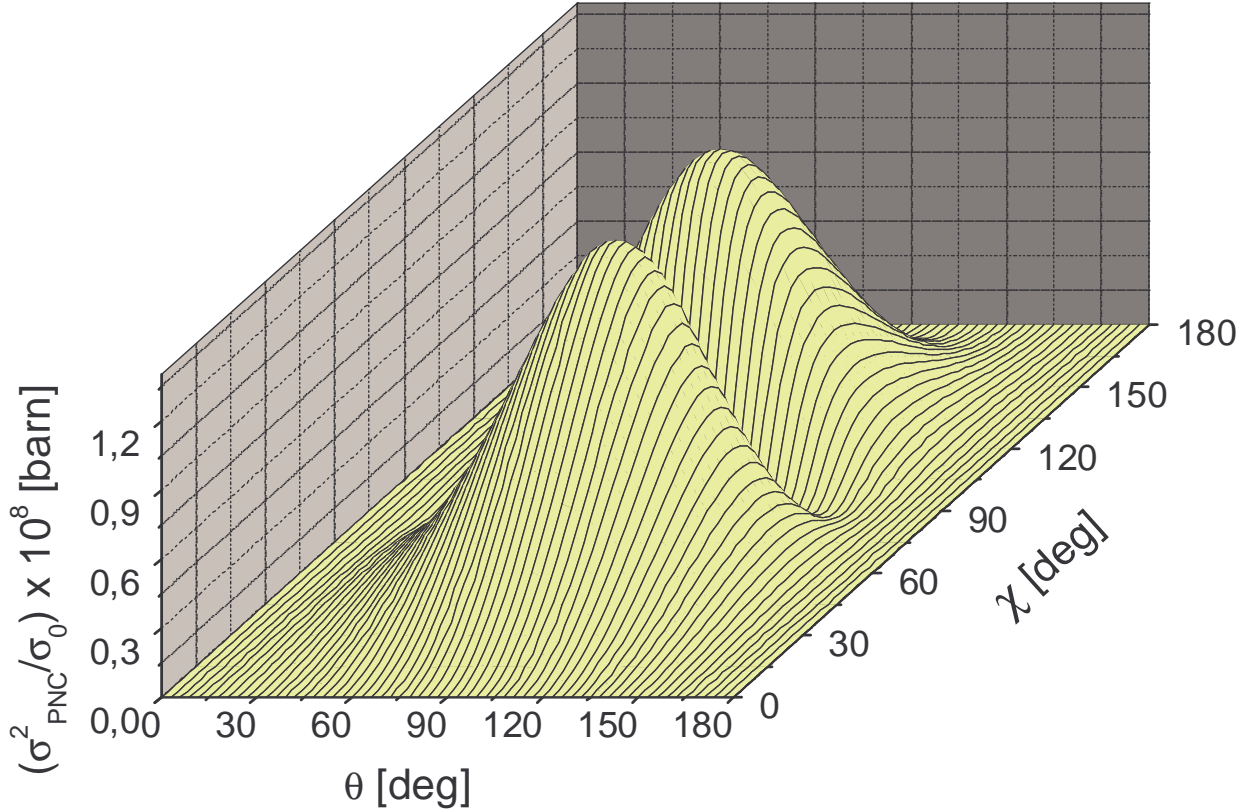} \caption{ The value $\sigma_{\rm
PNC}^2/\sigma_0 \sim 1/L_0$ as a function of the inclination
($\theta$) and polarization ($\chi$) angles for the radiative
recombination into the  $2^1S_0$ state of He-like thorium at the
incident electron energy of 1 eV. It is assumed that linearly
polarized photons are detected in the experiment with unpolarized
electrons and ions.  The PNC contribution $\sigma_{\rm PNC}$
changes the sign under the replacement $\chi \rightarrow
\pi-\chi$. } \label{Fig4}
\end{figure}

%
% ------------------------- Figure 5 ---------------------------- %
%
\begin{figure}
\includegraphics[width=\columnwidth]{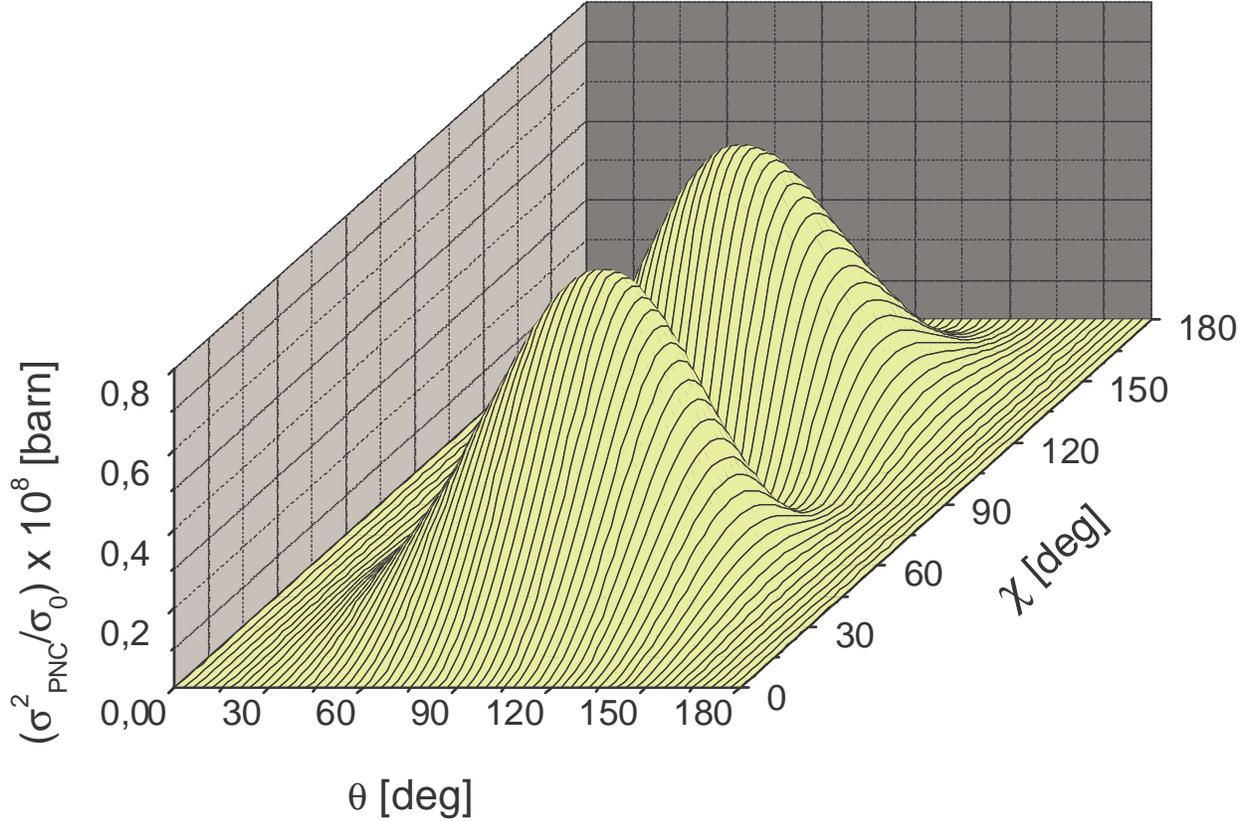} \caption{ The value $\sigma_{\rm
PNC}^2/\sigma_0 \sim 1/L_0$ as a function of the inclination
($\theta$) and polarization ($\chi$) angles for the radiative
recombination into the  $2^3P_0$ state of He-like thorium at the
incident electron energy of 1 eV. It is assumed that linearly
polarized photons are detected in the experiment with unpolarized
electrons and ions. The PNC contribution $\sigma_{\rm PNC}$
changes the sign under the replacement $\chi \rightarrow
\pi-\chi$. } \label{Fig5}
\end{figure}

The corresponding calculations have been also performed for the
$^{158}$Gd ion. It was found that, at the $2^3P_0$-$2^1S_0$ energy
splitting of 0.074 eV, the PNC effect is smaller than that for
thorium. In particular, in the first scenario for $p_i^0$ = 1 eV
the minimum luminosity for the RR into the $2^1S_0$ state amounts
to $2.5\times 10^{28}$ cm$^{-2}$s$^{-1}$ at $\theta$ = 0. The
corresponding values of the cross section contributions are
$\sigma_{0}$=287.65 barn and  $\sigma_{\rm PNC}$ = 0.0069 barn. In
the second scenario, at the same kinetic electron energy, the
minimum luminosity,  $L_0$ =  $7.1\times 10^{31}$
cm$^{-2}$s$^{-1}$, has been found at the angles $\theta$ = 92 and
$\chi$ = 76 with $\sigma_{0}$=2429.3 barn and  $\sigma_{\rm PNC}$
= 0.00038 barn. The maximum absolute value of the PNC
contribution, reached at $\theta$ = 91 and $\chi$ = 45, amounts to
$\sigma_{\rm PNC}$ = 0.00080 barn with $\sigma_{0}$=19488 barn.

\section{Conclusion}

In this paper, we investigated the PNC effect on the cross section
of the radiative recombination of an electron into the $2^1S_0$
and the  $2^3P_0$ state of heavy He-like ions. The calculations
were performed for the cases of thorium and gadolinium, where the
PNC effect is strongly enhanced due to near degeneracy of the
$2^1S_0$ and $2^3P_0$ states. Two scenarios of possible
experiments were studied. It was found that a promising situation
occurs in the scenario,  where linearly polarized photons are
detected in experiment with unpolarized electrons and ions and the
count rate difference between the two linear polarizations is
measured simultaneously at different azimuth angles.

\section{Acknowledgments}

This work was supported by DFG (Grant No. 436RUS113/950/0-1), by
RFBR (Grant No. 08-02-91967), and by GSI. The work of A.V.M.,
P.O.I, and V.M.S. is also supported by the Ministry of Education
and Science of Russian Federation (Program for Development of
Scientific Potential of High School, Grant No. 2.1.1/1136). V.M.S.
acknowledges financial support by the Alexander von Humboldt
Foundation. The work of A.V.M. was supported by the Dynasty
foundation. G.P. acknowledges financial support by DFG.

\newpage

\begin{table}
\footnotesize{ \caption{The $2^3P_0$ -  $2^1S_0$ energy difference
in He-like ions near the crossing points \cite{art05,kozh08a}, in
eV. } \vspace{0.2cm} \label{tab0}

\begin{tabular}{llllllll}
  \hline
  Eu ($Z$=63) &Gd ($Z$=64) &Tb ($Z$ =65) &Dy ($Z$ =66) &Ac
($Z$ =89) &Th ($Z$ =90) &Pa ($Z$ =91) &
U ($Z$ =92) \\
  \hline
  -0.249(69) &-0.023(74) & 0.29(12) & 0.462(84) & 1.52(40) & 0.44(40)
 & -0.43(50) & -2.81(8) \\
  \hline
\end{tabular}

%\begin{ruledtabular}
%\begin{tabular} {llllllll}
%\hline Eu ($Z$=63) &Gd ($Z$=64) &Tb ($Z$ =65) &Dy ($Z$ =66) &Ac
%($Z$ =89) &Th ($Z$ =90) &Pa ($Z$ =91) &
%U ($Z$ =92) \hspace*{0.115cm} \\
%\hline
% -0.249(69) &-0.023(74) & 0.29(12) & 0.462(84) & 1.52(40) & 0.44(40)
% & -0.43(50) & -2.81(8) \hspace*{0.115cm}
%\end{tabular}
%\end{ruledtabular}
}
\end{table}

\begin{table}
\footnotesize{ \caption{Numerical results for the radiative
recombination into the $2^1S_0$ and the $2^3P_0$  state of He-like
thorium at the inclination angles $\theta$ corresponding to the
minimum values of the luminosity $L_0$. The calculations are
performed for the scenario, where the electron beam is fully
polarized while the H-like ions are unpolarized and the photon
polarization is not detected. $L_0$ is the luminosity defined by
Eq. (\ref{lum}) at $T$ = 2 weeks, $\sigma_0  = (\sigma_+
+\sigma_-)/2$ is the cross section without the PNC effect, and
$\sigma_{\rm PNC} = (\sigma_+ -\sigma_-)/2$ is the PNC
contribution. } \vspace{0.2cm} \label{tab1}
%\begin{ruledtabular}
\begin{tabular} {c|cccc|cccc}
\hline
  &\multicolumn{4}{c}{RR into the $2^1S_0$ state }
                 & \multicolumn{4}{c}{RR into  the $2^3P_0$ state} \\
\hline $p_i^0$ [keV] &$\theta$ [grad] & $L_0$ [cm$^{-2}$s$^{-1}$]
& $\sigma_0$ [barn] & $\sigma_{\rm PNC}$ [barn]
 &$\theta$ [grad]
& $L_0$ [cm$^{-2}$s$^{-1}$] & $\sigma_0$ [barn] & $\sigma_{\rm
PNC}$ [barn]
\\ \hline
0.001& 180 & 9.1$\times 10^{26}$ & 4281.8 & 0.14 & 143 &
2.7$\times 10^{27}$ & 52856 & -0.28
 \hspace*{0.115cm} \\
0.005& 0 & 4.5$\times 10^{27}$ & 530.59 & 0.022 & 142 & 1.3$\times
10^{28}$ & 10757 & -0.058
 \hspace*{0.115cm} \\
0.010& 0  & 8.9$\times 10^{27}$ & 265.22 & 0.011 & 142  &
2.6$\times 10^{28}$ & 5346.7 & -0.029
 \hspace*{0.115cm} \\
0.050& 0 & 4.3$\times 10^{28}$ & 52.963 &  0.0023 & 140 &
1.3$\times 10^{29}$ & 1096.3 &  -0.0060
\hspace*{0.115cm} \\
0.100& 0  & 8.3$\times 10^{28}$ &26.440  & 0.0011 & 139  &
2.4$\times 10^{29}$ &551.34  & -0.0031
\hspace*{0.115cm} \\
0.500& 0  & 3.8$\times 10^{29}$ & 5.2346 & 0.00024 & 135  &
1.0$\times 10^{30}$ & 111.97 & -0.00067
\hspace*{0.115cm} \\
1.000& 0  & 7.0$\times 10^{29}$ &  2.5881 & 0.00012 & 131  &
1.9$\times 10^{30}$ & 57.751 & -0.00036
\hspace*{0.115cm} \\
5.000& 0 & 3.0$\times 10^{30}$ & 0.47948 & 2.6$\times 10^{-5}$ &
118 & 6.2$\times 10^{30}$ & 11.598 & -8.8$\times 10^{-5}$
\hspace*{0.115cm} \\
20.000& 0  & 1.1$\times 10^{31}$ & 0.094911 & 6.0$\times 10^{-6}$
& 94  & 1.5$\times 10^{31}$ & 2.7367 & -2.7$\times 10^{-5}$
\hspace*{0.115cm} \\
\hline
\end{tabular}
%\end{ruledtabular}
}
\end{table}

\begin{table}
\caption{Numerical results for the radiative recombination into
the $2^1S_0$ state of He-like  thorium at angles $\theta$ and
$\chi$ corresponding to the minimum values of the luminosity
$L_0$. The calculations are performed for the scenario, where  the
linearly polarized photons are detected in the experiment with
unpolarized electrons and ions. $L_0$ is the luminosity defined by
Eq. (\ref{lum}) at $T$ = 2 weeks, $\sigma_0$ is the cross section
without the PNC effect, and $\sigma_{\rm PNC}$ is the PNC
contribution, which changes the sign under the replacement $\chi
\rightarrow \pi-\chi$.
 } \vspace{0.2cm}
\label{tab2}
%\begin{ruledtabular}
\begin{tabular} {cccccc}
\hline $p_i^0$ [keV] &$\theta$ [grad] & $\chi$ [grad]
& $L_0$ [cm$^{-2}$s$^{-1}$] & $\sigma_0$ [barn] & $\sigma_{\rm PNC}$ [barn]\\
\hline
0.001& 94 & 68 & 3.0$\times 10^{29}$ & 10858 & 0.012  \hspace*{0.115cm} \\
0.005& 94 & 68 & 1.5$\times 10^{30}$ &2171.3  & 0.0024  \hspace*{0.115cm} \\
0.010& 93 & 68 & 3.0$\times 10^{30}$ & 1087.2  & 0.0012  \hspace*{0.115cm} \\
0.050& 92 & 68 & 1.5$\times 10^{31}$ & 217.57  & 0.00025  \hspace*{0.115cm} \\
0.100& 91 & 68 & 2.9$\times 10^{31}$ & 108.82  & 0.00012  \hspace*{0.115cm} \\
0.500& 87 & 68 & 1.5$\times 10^{32}$ & 21.706 & 2.5$\times
10^{-5}$ \hspace*{0.115cm}\\
\hline
\end{tabular}
%\end{ruledtabular}
\end{table}

\begin{table}
\caption{Numerical results in case of the radiative recombination
into the $2^3P_0$ state of He-like  thorium at angles $\theta$ and
$\chi$ corresponding to the minimum values of the luminosity
$L_0$. The calculations are performed for the scenario, where  the
linearly polarized photons are detected in the experiment with
unpolarized electrons and ions. $L_0$ is the luminosity defined by
Eq. (\ref{lum}) at $T$ = 2 weeks, $\sigma_0$ is the cross section
without the PNC effect, and $\sigma_{\rm PNC}$ is the PNC
contribution, which changes the sign under the replacement $\chi
\rightarrow \pi-\chi$.
 } \vspace{0.2cm}
\label{tab4}
%\begin{ruledtabular}
\begin{tabular} {cccccc}
\hline $p_i^0$ [keV] &$\theta$ [grad] & $\chi$ [grad]
& $L_0$ [cm$^{-2}$s$^{-1}$] & $\sigma_0$ [barn] & $\sigma_{\rm PNC}$ [barn]\\
\hline
0.001& 93 & 60 & 5.5$\times 10^{29}$ & 31208 & -0.015  \hspace*{0.115cm} \\
0.005& 92 & 60 & 2.8$\times 10^{30}$ &6235.4  & -0.0031  \hspace*{0.115cm} \\
0.010& 92 & 60 & 5.5$\times 10^{30}$ & 3116.4  & -0.0015  \hspace*{0.115cm} \\
0.050& 91 & 60 & 2.7$\times 10^{31}$ & 621.76  & -0.00031  \hspace*{0.115cm} \\
0.100& 91 & 60 & 5.4$\times 10^{31}$ & 310.28  & -0.00015  \hspace*{0.115cm} \\
0.500& 88 & 60 & 2.6$\times 10^{32}$ & 61.426 & -3.1$\times 10^{-5}$ \hspace*{0.115cm} \\
\hline
\end{tabular}
%\end{ruledtabular}
\end{table}

\begin{table}
\caption{Numerical results  for the radiative recombination into
the $2^1S_0$ and the $2^3P_0$  state of He-like thorium at  angles
$\theta$ and $\chi$ corresponding to the maximum absolute values
of the PNC contribution  $\sigma_{\rm PNC}$. The calculations are
performed for the scenario, where  the linearly polarized photons
are detected in the experiment with unpolarized electrons and
ions. The  $\sigma_{\rm PNC}$ contribution changes the sign under
the replacement $\chi \rightarrow \pi-\chi$.
 } \vspace{0.2cm}
\label{tab5}
%\begin{ruledtabular}
\begin{tabular} {ccc|cc|cc}
\hline & &  &\multicolumn{2}{c}{RR into the $2^1S_0$ state }
                 & \multicolumn{2}{c}{RR into  the $2^3P_0$ state} \\
\hline $p_i^0$ [keV] &$\theta$ [grad] & $\chi$ [grad] & $\sigma_0$
[barn] & $\sigma_{\rm PNC}$ [barn]
 & $\sigma_0$ [barn] & $\sigma_{\rm PNC}$ [barn] \\
\hline 0.001& 93 & 45 & 34410 & 0.018
& 51554 &-0.018  \hspace*{0.115cm} \\
0.005& 93 & 45 & 6879.7 & 0.0035
&10310 &-0.0035 \hspace*{0.115cm} \\
0.010& 92 & 45 & 3446.8 & 0.0018
& 5153.0 & -0.0018 \hspace*{0.115cm} \\
0.050& 92 & 45 & 688.69 & 0.00035
&  1030.1& -0.00036 \hspace*{0.115cm} \\
0.100& 91 & 45 & 344.70 & 0.00018
& 514.60 & -0.00018 \hspace*{0.115cm} \\
0.500& 88 & 45 & 68.940 & 0.000036 & 102.30 & -0.000036
\hspace*{0.115cm} \\
\hline
\end{tabular}
%\end{ruledtabular}
\end{table}

%\section*{References}
%\numrefs{10}


\begin{thebibliography}{99}

\bibitem{khr91} Khriplovich~I~B 1991 \emph{Parity Nonconservation in Atomic
Phenomena} (Gordon and Breach, London)

\bibitem{khr04} Khriplovich~I~B 2004 \emph{Phys. Scr. T} \textbf{112} 52

\bibitem{gin04} Ginges~J~S~M and  Flambaum~V~V 2004 \emph{Phys. Rep.} \textbf{397} 63

\bibitem{wood97} Wood~C~S, Bennett~S~C, Cho~D, Masterson~B~P, Roberts~J~L,
Tanner~C~E and Wieman~C~E 1997 \emph{Science} \textbf{275} 1759

\bibitem{ben99} Bennett~S~C and Wieman~C~E 1999 \emph{Phys. Rev. Lett.} {\bf 82}
2484; {\bf 83} 889

\bibitem{der00} Derevianko~A 2000 \emph{Phys. Rev. Lett.} {\bf 85} 1618; 2001
\emph{Phys. Rev. A} {\bf 65} 012106

\bibitem{koz01} Kozlov~M~G, Porsev~S~G and Tupitsyn~I~I 2001 \emph{Phys. Rev.
Lett.} {\bf 86} 3260

\bibitem{sush01} Sushkov~O~P 2001 \emph{Phys. Rev. A} {\bf 63} 042504

\bibitem{joh01} Johnson~W~R, Bednyakov~I and Soff~G 2001 \emph{Phys. Rev. Lett.}
{\bf 87} 233001

\bibitem{dzu02}
Dzuba~V~A, Flambaum~V~V, Ginges~J~S~M 2002 \emph{Phys. Rev. D}
{\bf 66} 076013

\bibitem{kuch02} Kuchiev~M~Y 2002 \emph{J. Phys. B} {\bf 35} L503 ; Kuchiev~M~Y and
Flambaum~V~V 2002 \emph{Phys. Rev. Lett.} {\bf 89} 283002; 2003
\emph{J. Phys. B} {\bf 36} R191; Flambaum~V~V and Ginges~J~S~M
2005 \emph{Phys. Rev. A} {\bf 72} 052115

\bibitem{mil02} Milstein~A~I, Sushkov~O~P and Terekhov~I~S 2002 \emph{Phys. Rev.
Lett.} {\bf 89} 283003; 2003 \emph{Phys. Rev. A} {\bf 67} 062103

\bibitem{sap03} Sapirstein~J, Pachucki~K, Veitia~A and Cheng~K~T 2003 \emph{Phys.
Rev. A} {\bf 67} 052110

\bibitem{sha05} Shabaev~V~M, Pachucki~K, Tupitsyn~I~I and Yerokhin~V~A 2005
\emph{Phys. Rev. Lett.} {\bf 94} 213002; Shabaev~V~M,
Tupitsyn~I~I, Pachucki~K, Plunien~G and Yerokhin~V~A 2005
\emph{Phys. Rev. A} {\bf 72} 062105

\bibitem{por09} Porsev~S~G, Beloy~K and Derevianko~A 2009 \emph{Phys. Rev. Lett.
} {\bf 102} 181601

\bibitem{gor74}
Gorshkov~V~G and Labzowsky~L~N 1974 \emph{Zh. Eksp. Teor. Fiz.
Pis'ma} {\bf 19} 768 [1974 \emph{JETP Lett.} {\bf 19} 394 ]; 1975
\emph{Zh. Eksp. Teor. Fiz.} {\bf 69} 1141 [1975 \emph{Sov. Phys.
JETP} {\bf 42} 581]

\bibitem{opp91} von Oppen~G 1991 \emph{Z. Phys. D} {\bf21} 181

\bibitem{sch89} Sch\"afer~A, Soff~G, Indelicato~P, M\"uller~B and
Greiner~W 1989 \emph{Phys. Rev. A} {\bf 40} 7362

\bibitem{kar92} Karasiev~V~V, Labzowsky~L~N and Nefiodov~A~V 1992 \emph{Phys. Lett. A}
{\bf 172} 62

\bibitem{dun96} Dunford~R~W 1996 \emph{Phys. Rev. A} {\bf 54} 3820

\bibitem{lab01} Labzowsky~L~N, Nefiodov~A~V, Plunien~G, Soff~G, Marrus~R and
Liesen~D 2001 \emph{Phys. Rev. A} {\bf 63} 054105

\bibitem{zol97} Zolotarev~M and Budker~D 1997 \emph{Phys. Rev. Lett.} {\bf 78}
4717

\bibitem{pin93} Pindzola~M~S 1993 \emph{Phys. Rev. A} {\bf 47} 4856

\bibitem{gri05} Gribakin~G~F, Currel~F~J, Kozlov~M~G and Mikhailov~A~I 2005
\emph{Phys. Rev. A} {\bf 72} 032109; arXiv:physics/0504129

\bibitem{eichler_book_95} Eichler J and Meyerhof W 1995 \emph{Relativistic Atomic Collisions}
(San Diego, CA: Academic)

\bibitem{sha02} Shabaev~V~M 2002 \emph{Phys. Rep.} {\bf 356} 119

\bibitem{kla02} Klasnikov~A~E, Artemyev~A~N, Beier~T, Eichler~J, Shabaev~V~M and
Yerokhin~V~A 2002 \emph{Phys. Rev. A} \textbf{66} 042711

\bibitem{sur02} Surzhykov~A, Fritzsche~S and St\"ohlker~Th 2002 \emph{J. Phys. B}
\textbf{35} 3713

\bibitem{fritzsche05} Fritzsche~S, Indelicato~P and St\"o{}hlker~Th 2005
               \emph{J. Phys. B: At. Mol. Phys.} \textbf{B38} S707


\bibitem{eic07} Eichler~J and St\"ohlker~Th 2007 \emph{Phys. Rep.} \textbf{439} 1

\bibitem{mai09} Maiorova~A~V, Surzhykov~A, Tashenov~S, Shabaev~V~M, Fritzsche~S, Plunien~G and
St\"ohlker~Th 2009 \emph{J. Phys. B: At. Mol. Opt. Phys.} {\bf 42}
125003

\bibitem{yer00}  Yerokhin~V~A, Shabaev~V~M, Beier~T and Eichler~J 2000 \emph{Phys. Rev.
A} \textbf{62} 042712

\bibitem{sur08} Surzhykov~A, Jentschura~U~D, St\"ohlker~Th and Fritzsche~S
2008 \emph{Eur. Phys. J. D} {\bf 46} 27

\bibitem{sha00} Shabaev~V~M, Yerokhin~V~A, Beier~T and Eichler~J 2000 \emph{Phys. Rev.
A} \textbf{61} 052112

\bibitem{ros57} Rose~M~E 1957 \emph{Elementary Theory of Angular Momentum} (New
York: Wiley)

\bibitem{salvat} Salvat~F, Fernandez-Varea~J~M and Williamson~Jr~W 1995
\emph{Comput. Phys. Commun.} {\bf 90} 151

\bibitem{Stohlker03} St\"ohlker~Th, Banas~D, Beyer~H~F, Gumberidze~A, Kozhuharov~C,
Kanter~E, Krings~T, Lewoczko~W, Ma~X, Protic~D, Sierpowski~D,
Spillmann~U, Tachenov~S and Warczak~A 2003 \emph{Nucl. Instr. and
Meth. in Phys. Res. B} \textbf{205} 210

\bibitem{Tashenov06} Tachenov~S, St\"ohlker~Th, Banas~D, Beckert~K, Beller~P,
Beyer~H~F, Bosch~F, Fritzsche~S, Gumberidze~A, Hagmann~S,
Kozhuharov~C, Krings~T, Liesen~D, Nolden~F, Protic~D,
Sierpowski~D, Spillmann~U, Steck~M and Surzhykov A 2006
\emph{Phys. Rev. Lett.} \textbf{97} 223202

\bibitem{art05} Artemyev~A~N, Shabaev~V~M, Yerokhin~V~A, Plunien~G and Soff~G
2005 \emph{Phys. Rev. A} {\bf 71} 062104

\bibitem{kozh08a} Kozhedub~Y~S and Shabaev~V~M unpublished.

\bibitem{kozh08b} Kozhedub~Y~S, Andreev~O~V, Shabaev~V~M, Tupitsyn~I~I, Brandau~C,
Kozhuharov~C, Plunien~G and St\"ohlker~Th 2008 \emph{Phys. Rev. A}
{\bf 77} 032501

\bibitem{yer06} Yerokhin~V~A, Indelicato~P and Shabaev~V~M, 2006 \emph{Phys. Rev. Lett.}
{\bf 97} 253004

\bibitem{bra08} Brandau~C, Kozhuharov~C, Harman~Z, M\"uller~A, Schippers~S,
Kozhedub~Y~S, Bernhardt~D, B\"ohm~S, Jacobi~J, Schmidt~E~W,
Mokler~P~H, Bosch~F, Kluge~H-J, St\"ohlker~Th, Beckert~K,
Beller~P, Nolden~F, Steck~M, Gumberidze~A, Reuschl~R, Spillmann~U,
Currell~F~J, Tupitsyn~I~I, Shabaev~V~M, Jentschura~U~D,
Keitel~C~H, Wolf~A and Stachura~Z 2008 \emph{Phys. Rev. Lett.}
{\bf 100} 073201
%\endnumrefs

\end{thebibliography}
\end{document}